# Event Identification in Social Networks

Fattane Zarrinkalam[†‡] and Ebrahim Bagheri[†]

[†]Laboratory for Systems, Software and Semantics (LS3), Ryerson University, Toronto, Canada
[‡]Department of Computer Engineering, Ferdowsi University of Mashhad, Mashhad, Iran
fattane.zarrinkalam@stu.um.ac.ir

Social networks enable users to freely communicate with each other and share their recent news, ongoing activities or views about different topics. As a result, they can be seen as a potentially viable source of information to understand the current emerging topics/events. The ability to model emerging topics is a substantial step to monitor and summarize the information originating from social sources. Applying traditional methods for event detection which are often proposed for processing large, formal and structured documents, are less effective, due to the short length, noisiness and informality of the social posts. Recent event detection techniques address these challenges by exploiting the opportunities behind abundant information available in social networks. This article provides an overview of the state of the art in event detection from social networks.

## 1. Overview

With the emergence and the growing popularity of social networks such as Twitter and Facebook, many users extensively use these platforms to express their feelings and views about a wide variety of social events/topics as they happen, in real time, even before they are released in traditional news outlets [1]. This large amount of data produced by various social media services has recently attracted many researchers to analyze social posts to understand the current emerging topics/events. The ability to identify emerging topics is a substantial step towards monitoring and summarizing the information on social media and provides the potential for understanding and describing real-world events and improving the quality of higher level applications in the fields of traditional news detection, computational journalism [2] and urban monitoring [3], among others.

### 1.1. *Problem Definition*

The task of Topic Detection and Tracking (TDT) to provide the means for news monitoring from multiple sources in order to keep users updated about news and developments. One of the first forums was the Topic Detection and Tracking (TDT) Forum, held within TREC [4]. There has been a significant interest in TDT in the past for static documents in traditional media [5, 6, 7, 8]. However, recently the focus has moved to Social Network data sources. Most of these works use Twitter as their source of information, because the information that the users publish on Twitter are more publicly accessible compared to other social networks.

To address the task of detecting topics from social media streams, a stream is considered to be made of posts which are generated by users in social network (e.g. tweets in the case of Twitter). Each post, in addition to its text formed by a sequence of words/terms, includes a user id and a timestamp. Additionally, a time interval of interest and a desired update rate is provided. The expected output of topic detection algorithms is the detection of emerging topics/events. In TDT, a topic is "a seminal event or activity, along with all directly related to events and activities." [4] where event refers to a specific thing that happens at a certain time and place [9, 10, 11]. A topic can be represented either through the clustering of documents in the collection or by a set of most important terms or keywords that are selected.

### 1.2. *Challenges*

The works proposed within the scope of the TDT have proven to be relatively well-established for Topic Detection in traditional textual corpora such as news articles [11]. However, applying traditional methods for event detection in the social media context poses unique challenges due to the distinctive features of textual data in social media [12] such as Time Sensitivity, Short Length, Unstructured Phrases, and Abundant Information.

- **Time Sensitivity.** Different from traditional textual data, the text in social media has real-time nature. Besides communicating and sharing ideas with each other, users in social networks may publish their feelings and views about a wide variety of recent events several times daily [7, 13, 14]. Users may want to communicate instantly with friends about "What they are doing (Twitter)" or "What is on their mind" (Facebook).

- **Short Length.** Most of social media platforms restrict the length of posts. For example, Twitter allows users to post tweets that are no longer than 140 characters. Similarly, Picasa comments are limited to 512 characters, and personal status messages on Windows Live Messenger are restricted to 128 characters. Unlike standard text with lots of words and their resulting statistics, short messages consist of few phrases or sentences. They cannot provide sufficient context information for effective similarity measure, the basis of many text processing methods [12, 29, 57].

- **Unstructured Phrases.** In contrast with well-written, structured, and edited news releases, social posts might include large amounts of meaningless messages, polluted and informal content, irregular, and abbreviated words, large number of spelling and grammatical errors, and improper sentence structures and mixed languages. In addition, in social networks, the distribution of content quality has high variance: from very high-quality items to low-quality, sometimes abusive content, which negatively affect the performance of the detection algorithms [61, 12].

- **Abundant Information.** In addition to the content itself, social media in general exhibit a rich variety of information sharing tools. For example, Twitter allows users to utilize the "#" symbol, called hashtag, to mark keywords or topics in a Tweet; an image is usually associated with multiple labels which are characterized by different regions in the image; users are able to build connection with others (link information). Previous text analytics sources most often appear as <user, content> structure, while the text analytics in social media is able to derive data from various aspects, which include user, content, link, tag, timestamps and others [62, 63, 64].

In the following, we explain different methodologies that have been proposed in the state of the art to tackle challenges in social networks.

## 2. Background Literature

According to the availability of the information about events, event detection algorithms can be classified into specified and unspecified techniques [13]. The specified techniques rely on specific information and features that are known about the event, such as a venue, time, type, and description. On the other hand, when there are no prior information available about the event, unspecified event detection technique rely on the social media streams to detect the occurrence of a real-world event.

### 2.1. *Specified Event Detection*

Specified event detection aims at identifying known social events which are partially or fully specified with its content or metadata information such as location, time, and venue. For example, Sakaki et al. [14] have focused on monitoring tweets posted recently by users to detect earthquake or rainbow. They have used three types of features: the number of words (statistical), the keywords in a tweet message, and the words surrounding users queries (contextual), to train a classifier and classify tweets into positive or negative cases. To identify the location of the event a probabilistic spatiotemporal model is also built. They have evaluated their proposed approach in an earthquake-reporting system in Japan. The authors have found that the statistical features provided the best results, while a small improvement in performance has been achieved by the combination of the three features.

Popescu and Pennacchiotti [15] have proposed a framework to identify controversial events. This framework is based on the notion of a Twitter snapshot which consists of a target entity, a given period, and a set of tweets about the entity from the target period. Given a set of Twitter snapshots, the authors first assign a controversy score to each snapshot and then rank the snapshots according to the controversy score by considering a large number of features, such as linguistic, structural, sentiment, controversy and external features in their model. The authors have concluded that Hashtags are important semantic features to identify the topic of a tweet. Further, they have found that linguistic, structural, and sentiment features provide considerable effects for controversy detection.

Benson et al. [16] have proposed a model to identify a comprehensive list of musical events from Twitter based on artist–venue pairs. Their model is based on a Conditional Random Field (CRF) to extract the artist name and location of the event. The input features to CRF model include word shape; a set of regular expressions for common emoticons, time references, and venue types; a bag of words for artist names extracted from external source (e.g., Wikipedia); and a bag of words for city venue names. Lee and Sumiya [17] have proposed a geosocial local event detection system, to identify local festivals. They have collected Twitter geotagged data for a specific region and used k-means algorithm applied to the geographical coordinates of the collected data to divide them into several regions of interest (ROI). The authors have found that an increased user activity, i.e., moving inside or coming to an ROI, combined with an increased number of tweets provides strong indicator of local festivals.

Becker et al. [18] have used a combination of simple rules and query building strategies to identify planned events from Twitter. They have identified tweets related to an event by utilizing simple query building strategies that derive queries from the structured description of the event and its associated aspects (e.g., time and venue). To provide high-precision tweets, they have asked an annotator to label the results returned by each strategy, then they have employed term-frequency analysis and co-location techniques to improve recall to identify descriptive event terms and phrases, which are then used recursively to define new queries. Similarly, Becker et al. [19] have proposed centrality-based approaches to extract high-quality, relevant, and useful related tweets to an event. Their approach is based on the idea that the most topically central messages in a cluster are more likely to reflect key aspects of the event than other less central cluster messages.

## 2.2. *Unspecified Event Detection*

The real-time nature of social posts reflect events as they happen about emerging events, breaking news, and general topics that attract the attention of a large number of users. Therefore, these posts are useful for unknown event detection. Three main approaches have been studied in the literature for this purpose: topic-modeling, document-clustering and feature-clustering approaches [1]:

### 2.2.1. *Topic Modeling Methods*

Topic modeling methods such as LDA, assume that a document is a mixture of topics and implicitly use co-occurrence patterns of terms to extract sets of correlated terms as topics of a text corpus [20]. More recent approaches have extended LDA to provide support for temporality including the recent Topics over Time (ToT) model [21], which simultaneously captures term co-occurrences and locality of those patterns over time and is hence able to discover more event-specific topics.

The majority of existing topic models including LDA and TOT, focus on regular documents, such as research papers, consisting of a relatively small number of long and high quality documents. However, social posts are shorter and noisier than traditional documents. Users in social networks are not professional writers and use very diverse vocabulary, and there are many abbreviations and typos. Moreover, the online social media websites have a social network full of context information, such as user features and user-generated labels, which have been normally ignored by the existing topic models. As a result, they may not perform so well on social posts and might suffer from the sparsity problem [22, 23, 24, 25]. To address this problem, some works aggregate multiple short texts to create a single document and discover the topics by running LDA over this document [25, 26, 27]. For instance, Hong and Davison [30], have combined all the tweets from each user as one document and apply LDA to extract the document topic mixture, which represents the user interest. However, in social networks a small number of users usually account for a significant portion of the content. This makes the aggregation process less effective.

There are some recent works that deal with the sparsity problem by applying some restrictions to simplify the conventional topic models or develop novel topic models for short texts. For example, Zhao et al. [28] have proposed the Twitter-LDA model. It assumes that a single tweet contains only one topic, which differs from the standard LDA model. Xueqi et al. [29] have proposed biterm topic model (BTM), a novel topic model for short texts, by learning the topics by directly modeling the generation of word co-occurrence patterns (i.e. biterms) in the whole corpus. BTM is extended by Yan et al. [57] by incorporating the burstiness of biterms as prior knowledge for bursty topic modeling and proposed a new probabilistic model named Bursty Biterm Topic Model (BBTM) to discover bursty topics in microblogs.

It should be noted that applying such restrictions and the fact that the number of topics in LDA is assumed to be fixed can be considered strong assumptions for social network content because of the dynamic nature of social networks.

### 2.2.2. *Document Clustering Methods*

Document-clustering methods extract topics by clustering related documents and consider each resulting cluster as a topic. They mostly represent textual content of each document as a bag of words or n-grams using TF/IDF weighting schema and utilize cosine similarity measures to compute the co-occurrence of their words/n-grams [31, 32]. Document-clustering methods suffer from cluster fragmentation problems and since the similarity of two documents may be sensitive to noise, they perform much better on long and formal documents than social posts which are short, noisy and informal [33]. To address this problem, some works, in addition to textual information, take into account other rich attributes of social posts such as timestamps, publisher, location and hashtags [33, 35, 36]. These works typically differ in that they use different information and different measures to compute the semantic distance between documents.

For example, Dong et al. [34] have proposed a wavelet-based scheme to compute the pairwise similarity of tweets based on temporal, spatial, and textual features of tweets. Fang et al. [35] have clustered tweets by taking into consideration multi-relations between tweets measured using different features such as textual data, hashtags and timestamp. Petrovic et al. [31] have proposed a method to detect new events from a stream of Twitter posts. To make event detection feasible on web-scale corpora, the authors have proposed a constant time and space approach based on an adapted variant of locality sensitive hashing methods. The authors have found that ranking according to the number of users is better than ranking according to the number of tweets and considering entropy of the message reduces the amount of spam messages in the output. Becker et al. [39] have first proposed a method to identify real-world events using an a classical incremental clustering algorithm. Then, they have classified the clusters content into real-world events or non-events. These non-events includes Twitter-centric topics, which are trending activities in Twitter that do not reflect any real-world occurrences. They have trained the classifier on the variety of features including temporal, social, topical, and Twitter-centric features to decide whether the cluster (and its associated messages) contains real-world event.

In the context of breaking news detection from Twitter, Sankaranarayanan et al. [37] have proposed TwitterStand which is a news processing system for Twitter to capture tweets related to late breaking news that takes into account both textual similarity and temporal proximity. They have used a naive Bayes classifier to separate news from irrelevant information and an online clustering algorithm based on weighted term vector to cluster news. Further, they have used hashtags to reduce clustering errors. Similarly, Phuvipadawat and Murata [38] have presented a method for breaking news detection in Twitter. They first sample tweets using predefined search queries, and then group them together to form a news story. Similarity between posts is based on tf-idf with an increased weight for proper noun terms, hashtags, and usernames. They use a weighted combination of number of followers (reliability) and the number of retweeted messages (popularity) with a time adjustment for the freshness of the message to rank each cluster. New messages are included in a cluster if they are similar to the first post and to the top-k terms in that cluster.

2.2.3. *Feature Clustering Methods*

Feature clustering methods try to extract features of topics from documents. opics are then detected by clustering features based on their semantic relatedness. As one of the earlier work that focused on Twitter data, Cataldi et. al. [40] have constructed a co-occurrence graph of emerging terms selected based on both the frequency of their occurrence and the importance of the users. The authors have applied a graph-based method in order to extract emerging topics. Similarly, Long et al. [41] have constructed a co-occurrence graph by extracting topical words from daily posts. To extract events during a time period, they have applied a top-down hierarchical clustering algorithm over the co-occurrence graph. After detecting events in different time periods, they track changes of events in consecutive time periods and summarize an event by finding the most relevant posts to that event. The algorithm by Sayyadi et al. [42] builds a term cooccurrence graph, whose nodes are clustered using a community detection algorithm based on betweenness centrality. Additionally, topic description is enriched with the documents that are most relevant to the identified terms. Graphs of short phrases, rather than of single terms, connected by edges representing lexical inclusion or similarity have also been used.

There are also some works that utilize signal processing techniques for event detection from social networks. For instance, Weng et al. [43] have used wavelet analysis to discover events in Twitter streams. First, they have selected bursty words by representing each word as a frequency-based signal and measuring the bursty energy of each word using autocorrelation. Then, they build a graph whose nodes are bursty words and edges are cross-correlation between each pair of bursty words and used graph-partitioning techniques to discover events. Similarly, Cordeiro [44] has used wavelet analysis for event detection from Twitter. This author has constructed a wavelet signal for each hashtag, instead of words, over time by counting the hashtag mentions in each interval. Then, he has applied the continuous wavelet transformation to get a time-frequency representation of each signal and used peal analysis and local maxima detection techniques to detect an event within a given time interval. He et al. [6] have used Discrete Fourier Transform to classify the signal for each term based on its power and periodicity. Depending on the identified class, the distribution of appearance of a term in time is modeled using one or more Gaussians, and the KL-divergence between the distributions is then used to determine clusters.

In general, most of these works are based on terms and compute similarity between pairs of terms based on their co-occurrence patterns. Petkos et. al. [45] have argued that the algorithms that are only based on pairwise co-occurrence patterns cannot distinguish between topics which are specific to a given corpus. Therefore, they have proposed a soft frequent pattern mining approach to detect finer grained topics. Zarrinkalam et al. [46] have inferred fine grained users' topics of interest by viewing each topic as a conjunction of several concepts, instead of terms, and benefit from a graph clustering algorithms to extract temporally related concepts in a given time period. Further, they compute inter-concept similarity by customizing the concepts co-occurrences within a single tweet to an increased, yet semantic preserving context.

3. **Application Areas**

There are a set of interesting applications of event/topic detection systems and methods. Health monitoring and management is an application in which the detection of events plays an important role. For example, Culotta [49] have explored the possibility of tracking influenza by analyzing Twitter data. They have proposed an approach to predict influenza-like illnesses rates in a population to identify influenza-related messages and compare a number of regression models to correlate these messages with U.S. Centers for Disease Control and Prevention (CDC) statistics. Similarly, Aramaki [52] have identified flu outbreaks by analysing tweets about Influeza. Their results are similar to Google-trends based flu outbreak detection especially in the early stages of the outbreak

Paul and Dredze [50] have proposed a new topic model for Twitter, named Ailment Topic Aspect Model (ATAM), that associates symptoms, treatments and general words with diseases. It produces more detailed ailment symptoms and tracks disease rates consistent with published government statistics (influenza surveillance) despite the lack of supervised influenza training data. In [51], the authors have used Twitter to identify posts which are about health issues and they have investigated what types of links the users consult for publishing health related information.

Natural events detection (Disasters) is another application for the automatic detection of events from social network. For example, Sakaki et al. [14] have proposed an algorithm to monitor the real-time interaction of events, such as earthquakes in Twitter. Their approach can detect an earthquake with high probability by monitoring tweets and detects earthquakes promptly and sends e-mails to registered users. The response time of the system is shown to be quite fast, similar to the Japan Meteorological Agency. Cheong and Cheong [53] have analysed the tweets during Australian floods of 2011 to identify active players and their effectiveness in disseminating critical information. As their secondary goal, they have identified the most important users among Australian floods to be: local authorities (Queensland Police Services), political personalities (Premier, Prime Minister, Opposition Leader and Member of Parliament), social media volunteers, traditional media reporters, and people from not for profit, humanitarian, and community associations. In [54], the authors have applied visual analytics approach to a set of georeferenced Tweets to detect flood events in Germany providing visual information on the map. Their results confirmed the potential of Twitter as a distributed "social sensor". To overcome some caveats in interpreting immediate results, they have explored incorporating evidence from other data sources.

Some applications with marketing purpose have also utilized event detection methods. For example, Medvent et al. [55] have focused on detecting events related to three major brands including Google, Microsoft and Apple. Examples of such events are the release of a new product like the new iPad or Microsoft Security Essential software. In order to achieve the desired outcome, the authors study the sentiment of the tweets. Si et al. [56] have proposed a continuous Dirichlet Process Mixture model for Twitter sentiment, to help predict the stock market. They extract the sentiment of each tweet based on its opinion words distribution to build a sentiment time series. Then, they regress the stock index and the Twitter sentiment time series to predict the market.

There are also some works that model user's interests over detected events from social networks. For example, Zarrinkalam et al. [47] have proposed a graph-based link prediction schema to model a user's interest profile over a set of topics/events present in Twitter in a specified time interval. They have considered both explicit and implicit interests of the user. Their approach is independent of the underlying topic detection method, therefore, they have adopted two types of topic extraction methods: feature clustering and LDA approaches. Fani et al. [48] have proposed a graph-based framework that utilizes multivariate time series analysis to tackle the problem of detecting time-sensitive topic-based communities of user who have similar temporal tendency with regards to topics of interests in Twitter. To discover topics of interest from Twitter, they have utilized an LDA-based topic model that jointly captures word co-occurrences and locality of those patterns over time.

**4. Conclusion and Future Directions**

Due to the fast-growing and availability of social network data, many researchers has recently become attracted to event detection from social networks. Event detection aims at finding real-world occurrences that unfold over space and time. The problem of event detection from social networks has faced different challenges due to the short length, noisiness and informality of the social posts. In this article we presented an overview of the recent techniques to address this problem. These techniques are classified according to the type of target event into specified or unspecified event detection. Further, we provided some potential applications in which event detection techniques are utilized.

While there are many works related to event detection from social networks, one challenge that has to be addressed in this research area is the lack of public datasets. Privacy issues along with Social Network companies' terms of use hinder the availability of shared data. This obstacle is of great significance since it relates to the repeatability of experiments and comparison between approaches. As a result, most of the current approaches have focused on a single data source, specially the Twitter platform because of the usability and accessibility of the Twitter API. However, being dependent on a single data source entails many risks. Therefore, one future direction can be monitoring and analyzing the events and activities from different social network services simultaneously. As an example, Kaleel et al. [58] have followed this idea and utilized Twitter posts and Facebook messages for event detection. They have used LSH to classify messages. The proposed algorithm first independently identifies new events (first stories) from both sources (Twitter, Facebook) and then hashes them into clusters.

As another future direction, there is no method in the field of event detection from social networks which is able to automatically answer the following questions for each detected event: what, when, where, and by whom. Therefore, improving current methods to address these questions can be a new future direction. As a social post is often associated with spatial and temporal information, it is possible to detect when and where an event happens.

Several further directions can be explored to achieve efficient and reliable event detection systems such as: investigating how to model the social streams together with other data sources, like news streams to better detect and represent events [60], designing better feature extraction and query generation techniques, designing more accurate filtering and detection algorithms as well as techniques to support multiple languages [59].